\documentclass[fleqn,12pt,twoside]{article}
\usepackage{espcrc1}
\usepackage{graphicx}
\usepackage{txfonts}
\usepackage{aasjournals}

\newcommand{\unitspace}{\ensuremath{\,}}
\newcommand{\usp}{\unitspace}
\newcommand{\numberspace}{\ensuremath{\;}}
\newcommand{\nsp}{\numberspace}
\newcommand{\unitstyle}[1]{\ensuremath{\mathrm{#1}}}

\newcommand{\kilo}{\unitstyle{k}}

\newcommand{\cm}{\unitstyle{cm}}

\newcommand{\meter}{\unitstyle{m}}

\newcommand{\second}{\unitstyle{s}}

\newcommand{\Kelvin}{\unitstyle{K}}
\newcommand{\K}{\Kelvin}  %degrees Kelvin

\newcommand{\km}{\kilo\meter}   %kilometers
\newcommand{\code}[1]{\textsc{#1}}
\newcommand{\Flash}{\code{Flash}}

\newcommand{\nuclei}[2]{\ensuremath{\mathrm{^{#1}#2}}}
\newcommand{\C}{\nuclei{12}{C}}
\newcommand{\Ox}{\nuclei{16}{O}}
\newcommand{\Mg}{\nuclei{24}{Mg}}
\newcommand{\Si}{\nuclei{28}{Si}}
\newcommand{\Ni}{\nuclei{56}{Ni}}

\newcommand{\adndt}{At.~Data~Nucl.~Data~Tables}
\title{Type Ia Supernovae: Simulations and Nucleosynthesis}

\author{E. F. Brown\address{Department of Physics and Astronomy and the
    Joint Institute for Nuclear Astrophysics, Michigan State University,
    East Lansing, MI 48824 USA}\thanks{This work is supported in part by
    the U.S. Department of Energy under Grant No. B523820 to the Center
    for Astrophysical Thermonuclear Flashes at the University of
    Chicago.},
  A. C. Calder\address[flash]{Center for Astrophysical Thermonuclear
    Flashes, The University of Chicago, Chicago, IL 60637 USA},
  T. Plewa\address{Nicolaus Copernicus Astronomical Center, Bartycka 18, 
    00716 Warsaw, Poland}\thanks{Current address: Center
    for Astronomical Thermonuclear Flashes, The University
    of Chicago, Chicago, IL 60637 USA},
  P. M. Ricker\address{Department of Astronomy, University of Illinois,
    Urbana, IL  61801 USA}\thanks{National Center for Supercomputing
    Applications, Urbana, IL  61801 USA}
  K. Robinson\addressmark[flash],
  J. B. Gallagher\addressmark[flash]
}

\begin{document}

% typeset front matter
\maketitle

\begin{abstract}
   We present our first nucleosynthesis results from a numerical
   simulation of the thermonuclear disruption of a static cold
   Chandrasekhar-mass \C/\Ox\ white dwarf.  The two-dimensional
   simulation was performed with an adaptive-mesh Eulerian hydrodynamics
   code, \Flash, that uses as a flame capturing scheme the evolution of
   a passive scaler.
   To compute the isotopic yields and their velocity distribution,
   10,000 massless tracer particles are embedded in the star. 
   The particles are advected along streamlines and provide a Lagrangian
   description of the explosion.  We briefly describe our verification
   tests and preliminary results from post-processing the particle
   trajectories in $(\rho, T)$ with a modest (214 isotopes) reaction
   network.
\end{abstract}

\section{Introduction}

Type Ia supernovae are bright stellar explosions that are distinguished
from other supernovae by a lack of hydrogen in their spectra.  A widely
accepted progenitor for these events is a close binary system in which
the primary component is a degenerate C-O rich white dwarf that accretes
from a low-mass companion.  The explosion is believed to ignite in the
center of the white dwarf when its center becomes sufficiently dense and
hot.  For a review of different explosion models, see
\cite{hillebrandt.niemeyer:SNeReview} and references therein.

Constraints on the nature of the explosion come from studies of the
supernova light curve and the composition of the ejecta.
Multi-dimensional hydrodynamical simulations cannot include more than a
few isotopes and are typically Eulerian.  Here we report on our
implementation of tracer particles into a simulation of a deflagrating
C-O white dwarf.  These particles are advected along streamlines and
provide a Lagrangian description of the explosion that can be
post-processed with a realistic reaction network.  We discuss our
verification tests, a 2-D deflagration, and the resulting
nucleosynthetic yields.

\section{Simulation with Tracer Particles} 
\label{sec:simulation}

The simulations of the deflagrating white dwarf were performed with the
\Flash\ code~\cite{fryxell+00}. \Flash\ solves the Euler equations for
compressible flow and the Poisson equation for self-gravity on an
adaptive mesh.  These simulations employ a flame capturing scheme (see
\cite{khokhlov95}) that advances the flame by evolving a passive scalar
variable with an advection-reaction-diffusion equation.  The flame model
assumes that the flame propagates at a given speed $v =
\max(v_\mathrm{lam}, v_\mathrm{turb})$, where $v_\mathrm{lam}$ is the
laminar flame speed~\cite{timmes+92}, and $v_\mathrm{turb}$ is a
turbulent flame speed based on the assumption that the turbulent burning
on macroscopic scales is driven by Rayleigh-Taylor
instability~\cite{khokhlov95}.  The scalar variable is used to model the
burning from \C\ to \Mg.  Further burning to Si-group and relaxation to
NSE (here taken to be \Ni) are modeled as two consecutive stages of
burning following the burn from C to Mg.

In addition to the flame model, the simulation of the deflagration had
10,000 tracer particles added to track the evolution of the fluids.  The
technique of using tracer particles in an Eulerian scheme has been used
in simulations of core-collapse supernovae
\cite{nagataki.hashimoto.ea:explosive,travaglio.kifonidis.ea:nucleosynthesis}
and Type Ia supernovae \cite{travaglio.hillebrandt.ea:nucleosynthesis}.
The initial spatial distribution of the particles follows the mass
distribution. The tracer particles were evolved by a second-order
predictor-corrector method~\cite{rider+95}, with the velocity at the
particle position computed by parabolic interpolation.

\subsection{Verification}
\label{sec:verification}

We have partially verified the tracer particle module on a smooth
hydrodynamic problem, the isentropic vortex.  This test problem has a
known analytic solution and is suitable for solution verification
studies (see \cite{calder02}, and references therein). We used a uniform
grid of 128$\times$128 equidistant zones and 10,000 randomly distributed
tracer particles. In our analysis we focused on a subset of three test
particles located at $r=0.26$, 0.767, and 3.873\usp\cm. We evolved this
system for $t = 100\usp\second$, which corresponds to 21.3 orbital
periods for the inner particle and about 15.4 orbital periods for the
middle particle. The outer particle moves very slowly and traveled a
negligible fraction of its orbit over the simulation.

\begin{figure}[htbp]
  \centering
  \includegraphics[width=80mm]{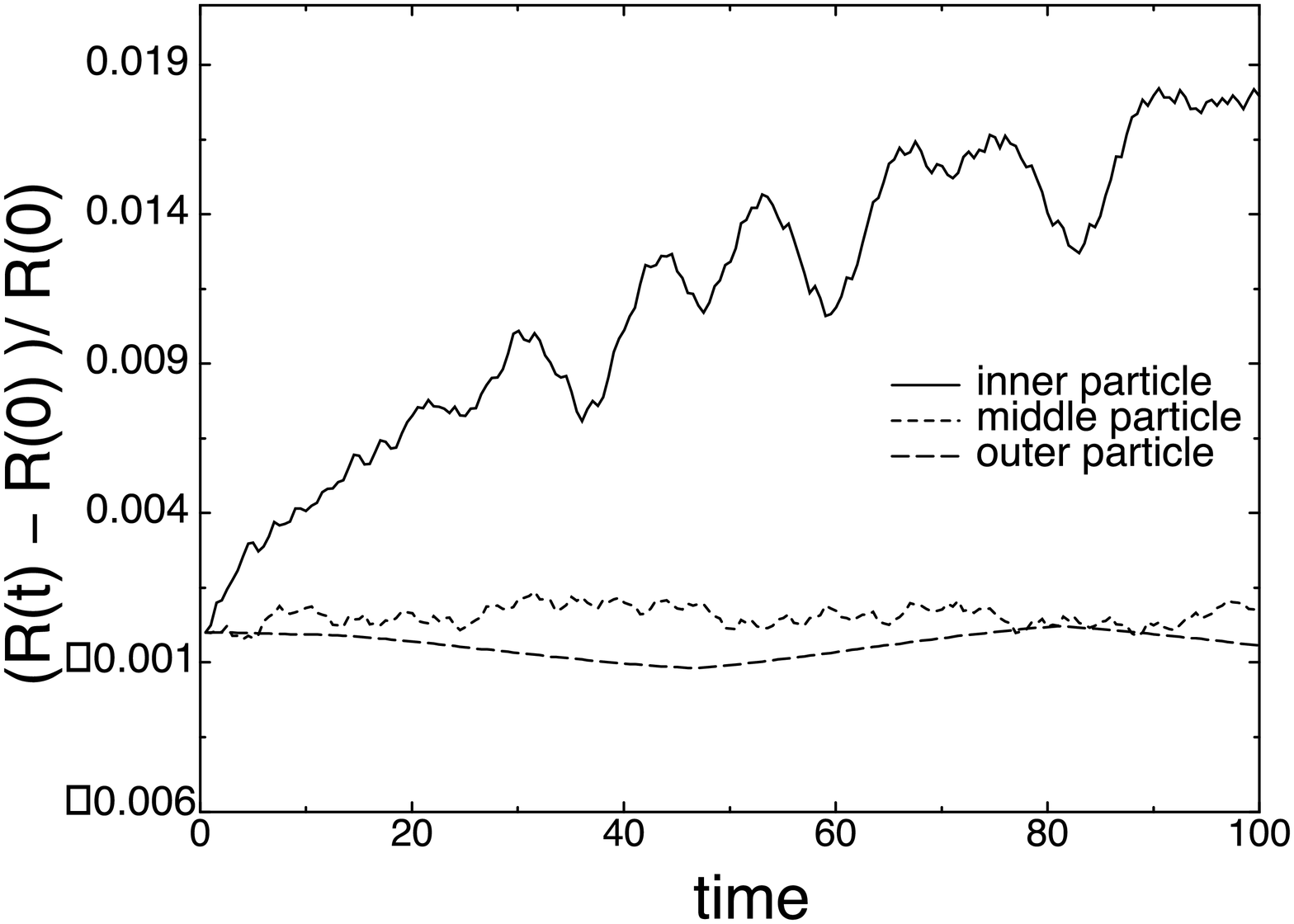}\hfill
  \includegraphics[width=80mm]{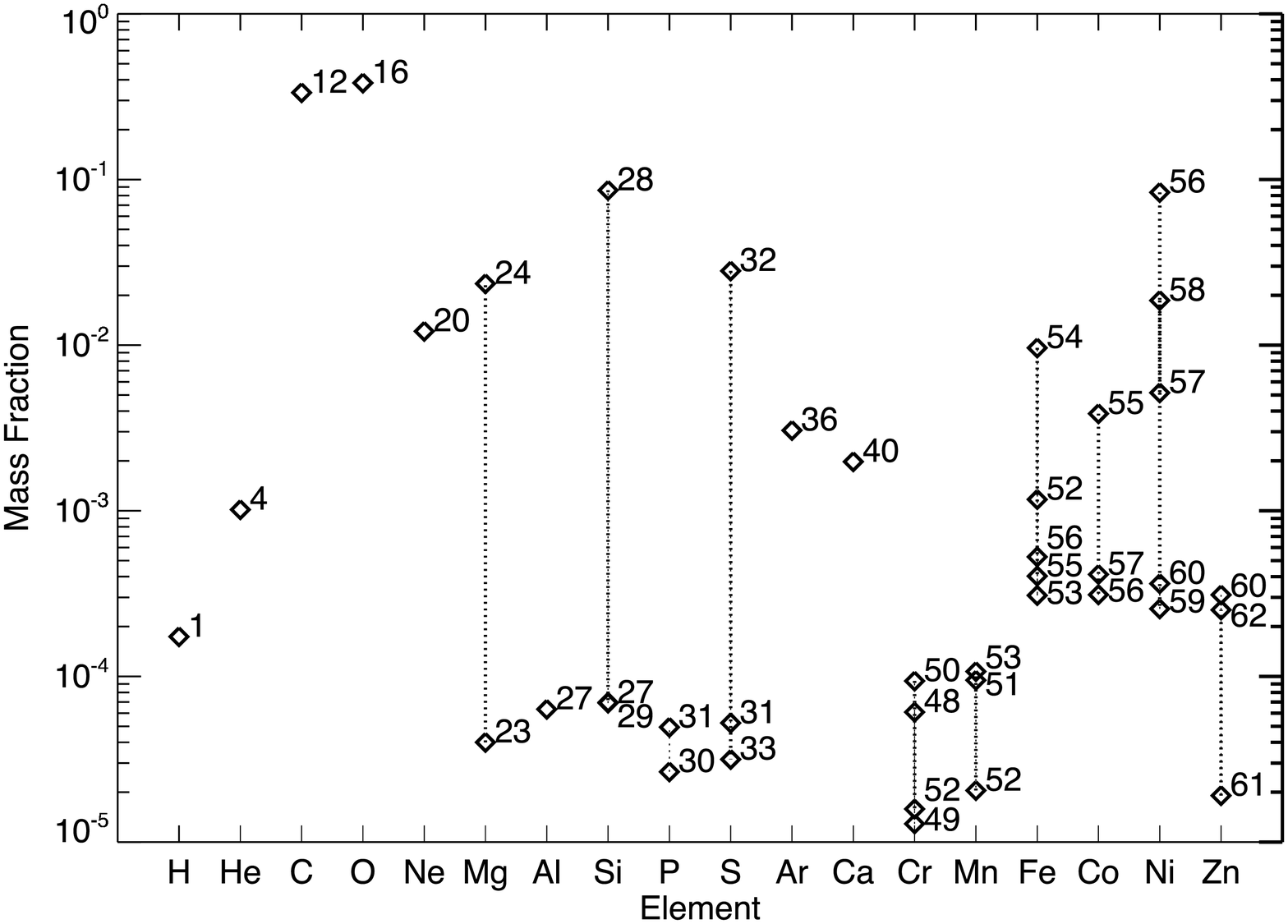}
  \caption{\emph{Left panel} Relative error in the position of test tracer
    particles in the isentropic vortex verification test problem.
    \emph{Right panel} Mass fractions of the isotopes synthesized by the
    deflagration.  Dotted lines connect isotopes of the same element.
    The number beside each point refers to the mass number of that
    isotope.}
  \label{fig:all}
\end{figure}

Fig.~\ref{fig:all} (\emph{left panel}) shows the relative error in the
particle radius at the final time.  The deviation from the nominal orbit
is generally small and is less than 1\% for $t < 20\usp\second$, or
about 4 orbits for the inner particle.  At later times the error
steadily grows but appears to be increasing for only the inner particle.
The orbit of the inner particle is tight (resolved by about only 20
zones), and the general increase in radius likely results from diffusion
in the hydrodynamic solver in addition to the intrinsic error in tracing
the flow.  While separating these effects is a future endeavor, we are
confident that our implementation is correct and that tracer particles
closely follow the streamlines.

\subsection{A 2-D Deflagration}
\label{sec:deflagration}

The two-dimensional deflagration simulation was performed in cylindrical
coordinates and azimuthal symmetry on a domain with $r = [0,8
192]\nsp\kilo\meter$ and $z=[-8 192, 8 192]\nsp\kilo\meter$ and an
effective resolution of 8\nsp\kilo\meter.  The initial model was an
isothermal white dwarf of temperature $T_\mathrm{wd}=5\times10^7\nsp\K$,
mass $M_\mathrm{wd}=1.3\nsp M_\odot$, and radius $R_\mathrm{wd}=2
130\nsp\kilo\meter$.  The outer boundary conditions were hydrostatic
outflow-only and reflecting boundary conditions were imposed along the
symmetry axis.  The deflagration was ignited by making the innermost
100\nsp\km\ the white dwarf into completely burned material in
hydrostatic equilibrium with its surroundings.  The flame was then
evolved by integrating the hydrodynamic equations supplemented by the
advection-diffusion-reaction model described above.  The deflagration
was evolved for 2.0\nsp\second, by which time $\approx 0.4\nsp M_\odot$
had burned.

\section{Nucleosynthesis}
\label{sec:nucleosynthesis}

The isotopic abundances were calculated by integrating a reaction
network along the trajectories in $(\rho,T)$ space followed by the
tracers.  We use a network consisting of 214 isotopes ranging from
neutrons, protons, and $\alpha$-particles to \nuclei{70}{Zn}.  The
coupled ODEs are integrated using an adaptive semi-implicit scheme with
a sparse matrix solver \cite{timmes:integration}.  We use the reaction
rate compilation \code{reaclib} \cite{rauscher.thielemann:rates}, and
incorporate the tabulated weak rates from \cite{langange.weak}.
Screening is computed according to the prescription of
\cite{graboske.dewitt.ea:screening}.  Since each particle is
independent, the calculation is embarrassingly parallel, and large
numbers of particles can be quickly post-processed on a multi-CPU
machine.

Fig.~\ref{fig:all} (\emph{right panel}) shows the mass fractions of the
isotopes synthesized in the deflagration.  One notes that most of the
original C-O did not burn and that roughly equal masses of \Ni\ and \Si\ 
were made.  Only about $0.15\nsp M_\odot$ of Ni-peak elements are
synthesized.  This is much less than that found in one-dimensional
calculations such as W7 \cite{iwamoto.brachwitz.ea:nucleosynthesis} and
recent 3-D octant calculations
\cite{travaglio.hillebrandt.ea:nucleosynthesis}.  For our chosen
configuration, the burning is incomplete.  Large plumes form and rise
toward the surface of the white dwarf.  The large scale circulation
arising from these flows tends to homogenize the composition as a
function of radius: unburned \C\ and \Ox, intermediate mass elements and
Ni-peak elements are present at roughly constant abundances from the
center to about 0.9 of the WD mass (as far as the rising plumes had
risen for this simulation).

We did repeat the calculations with a starting abundance that included
\nuclei{22}{Ne}.  From charge and mass conservation, one can show that
the mass of \Ni\ should decrease linearly with the initial mass fraction
of \nuclei{22}{Ne} \cite{timmes.brown.ea:variations}.  Calculations with
an initial \nuclei{22}{Ne} mass fraction of 0.02 and 0.06 also find a
linear dependence of the mass fraction of \Ni, but with a 10\% steeper
slope.  Because the underlying hydrodynamical model was the same,
however, we cannot draw further conclusions about the relationship
between ejected \Ni\ mass and progenitor metallicity.  We also checked
how the computed abundances scaled with the number of tracer particles.
For a subset of 5000 randomly selected particles, the difference in the
\Ni\ mass from that computed with 10,000 particles is 8\%.

We finally note that although only about 10\% of the star is burnt to
Fe-peak elements of that fraction, almost all of it is in the form of
\Ni; the mass fraction of \nuclei{58}{Ni}, \nuclei{54}{Fe} and
\nuclei{56}{Fe} are 0.019, 0.009, and 0.0005, respectively.  This is in
contrast to 1-D models, for which electron captures convert the
innermost $\approx 0.2\nsp M_\odot$ to \nuclei{56}{Fe}
\cite{brachwitz:ec}.  The rising plumes of hot ash carry the burnt
material to lower density, where electron capture rates are lower.  The
effect of rising bubbles of hot fuel to mix the white dwarf and leave
masses of C-O fuel unburnt has been noted by others (for example,
\cite{gamezo+03}) and indeed, is a motivation for suggesting a
subsequent transition to detonation.

In conclusion, we have implemented particle tracers in an Eulerian,
adaptive mesh hydrodynamics code, verified that they closely followed
streamlines, and used them for post-processing a 2-D simulation of a
deflagrating C-O white dwarf.  Future 3-D simulations will use this
capability for comparison against other simulations
\cite{travaglio.hillebrandt.ea:nucleosynthesis} and observations.


\begin{thebibliography}{99}

\bibitem{hillebrandt.niemeyer:SNeReview}
   W. Hillebrandt and J.~C. Niemeyer, \araa\ 38 (2000) 191

\bibitem{fryxell+00}
B. Fryxell, et~al., \apjs\ 131 (2000) 273

\bibitem{khokhlov95}
A.~M. Khokhlov, \apj\ 449 (1995) 695

\bibitem{timmes+92}
F.~X. Timmes and S.~E. Woosley, \apj\ 396 (1992) 649

\bibitem{nagataki.hashimoto.ea:explosive}
  S. Nagataki, M. Hashimoto, K. Sato, and S. Yamada, \apj\ 486 (1997) 1026

\bibitem{travaglio.kifonidis.ea:nucleosynthesis}
  C. Travaglio, K. Kifonidis, and E. M{\" u}ller, New Astro.
  Rev. 48 (2004) 25

\bibitem{travaglio.hillebrandt.ea:nucleosynthesis}
  C. Travaglio, W. Hillebrandt, M. Reinecke, and F.-K. Thielemann, \aap\ 425 (2004) 1029

\bibitem{rider+95}
W. J. Rider and D. B. Kothe {\em A Marker Particle Method
for Interface Tracking} LA-UR-95-1740 (1995)

\bibitem{calder02}
  A.~C. Calder, et~al., \apjs\ 143 (2002) 201

\bibitem{timmes:integration}
  F. X. Timmes, \apjs\ 124 (1999) 241

\bibitem{rauscher.thielemann:rates}
  T. Rauscher and F.-K. Thielemann, \adndt 75 (2000) 1

\bibitem{langange.weak}
  K. Langanke and G. Mart\'inez-Pinedo, \adndt\ 79 (2001) 1

\bibitem{graboske.dewitt.ea:screening}
  H.~C. Graboske, H.~E. Dewitt, A.~S. Grossman, and M.~S. Cooper, \apj\
  181 (1973) 457

\bibitem{iwamoto.brachwitz.ea:nucleosynthesis}
  K. Iwamoto, F. Brachwitz, K. Nomoto, N. Kishimoto, H. Umeda,
  W.~R. Hix, and F.-K. Thielemann, \apjs\ 125 (1999) 439

\bibitem{timmes.brown.ea:variations}
  F.~X. Timmes, E.~F. Brown and J.~W. Truran, \apj\ 590 (2003) L83

\bibitem{brachwitz:ec}
  F. Brachwitz~et al., \apj\ 536 (2000) 934

\bibitem{gamezo+03}
V. Gamezo, et~al., Science 299 (2003) 77


\end{thebibliography}
\end{document}